\documentclass[letterpaper,twocolumn,10pt]{article}
\usepackage{paperstyle}

\usepackage{amsmath}
\usepackage{enumitem}
\usepackage{xspace}
\usepackage{graphicx}
\usepackage{tikz}
\usepackage{cleveref}
\hypersetup{
    colorlinks=true,
    linkcolor=blue,
    filecolor=magenta,
    urlcolor=cyan,
    pdfpagemode=FullScreen
}
\usepackage{soul}
\usepackage{amssymb}
\usepackage{pifont}
\usepackage[framemethod=TikZ]{mdframed}
\usepackage{booktabs}
\usepackage{algorithm}
\usepackage{algorithmic}
\newcommand{\mypara}[1]{\vspace{2pt}\noindent\textbf{{#1. }}}

\begin{document}

\newcommand{\name}{JITterFlip\xspace}
\date{}
\title{\name: Uncovering Fault Attack Surfaces in JIT-Compiled LLM Serving}
\author{
Tairui Wang$^{1}$, Zhi Zhang$^{2}$, Yansong Gao$^{2}$, Xin Zhang$^{3}$,\\
Qingni Shen$^{1}$, and Zhonghai Wu$^{1}$\\[3pt]
\small $^{1}$Peking University \quad $^{2}$The University of Western Australia \quad $^{3}$Shandong University\\
\small \texttt{tairui.wang@stu.pku.edu.cn}, \texttt{qingnishen@pku.edu.cn}, \texttt{wuzh@pku.edu.cn}\\
\small \texttt{zzhangphd@gmail.com}, \texttt{gao.yansong@hotmail.com}, \texttt{zhangxin00sdu@gmail.com}
}
\maketitle

\begin{abstract}
LLMs are widely deployed through cloud-hosted inference services, where Just-in-Time (JIT) compilation is used to reduce recurring framework and GPU-launch overhead. JIT serving introduces a host-side control plane that selects compiled artifacts and orchestrates their execution on the GPU.
Meanwhile, the shared cloud setting has motivated a growing body of bit-flip attacks (BFAs) against LLM/DNN inference.
Most existing BFAs target model parameters/weights require model-specific knowledge. A smaller body of work reduces this dependency by faulting executable code, yet still corrupts code that directly implements model computation, limiting their attack effect to {inference depletion}.

We present \name, the first BFA targeting the \emph{host-side JIT serving control plane} of GPU-based LLM inference. By faulting CPU-resident serving decisions rather than model computation, \name enables both \emph{gibberish output generation} and a \emph{correct-output sponge attack}. To identify exploitable targets in a large JIT compiler stack, \name develops a decision-guided fault-vulnerable code analysis.

Across four text and multimodal LLM workloads, the identified vulnerable code faults exhibit cross-model transferability, produce gibberish outputs with PPL ratios of \(15.45\times\) to \(2.48{\times}10^{6}\times\), and demonstrate correct-output sponge attacks with latency amplification of \(2.03\times\) to \(181.90\times\). \name also bypasses recent BFA defenses for LLMs while retaining both attack effects. Last, we demonstrate end-to-end Rowhammer attacks across four LLMs: a single bit flip in CPU-resident branch code propagates across the CPU--GPU boundary to disrupt GPU-executed inference without direct access to GPU memory, reaching up to \(7.23{\times}10^{6}\times\) PPL amplification or \(124.97\times\) latency amplification while preserving the exact generated output.

\end{abstract}

\section{Introduction}
Large language models (LLMs) have become foundational to modern AI systems, supporting a broad range of text, code, reasoning, document-analysis, and multimodal applications~\cite{zhao2023survey,singhal2023large}. Their rapid adoption has also driven increasing demand for efficient inference serving.
Just-in-Time (JIT) compilation is particularly attractive for LLM serving because autoregressive generation repeatedly exercises the model execution path. It significantly reduces the recurring framework and GPU-launch overhead by reusing optimized compiled execution, while retaining runtime specialization for changing serving conditions. This execution mode has received mainstream support. For example, vLLM V1 enables PyTorch \texttt{torch.compile} by default and combines compiled-code caching, shape specialization, and CUDA-Graph execution to accelerate LLM serving~\cite{vllmTorchCompile2025,vllmCompilationConfig2026,vllmCudaGraphs2026}.

Meanwhile, LLMs are widely served through cloud-hosted GPU/CPU platforms and model-as-a-service APIs, where mutually untrusted tenants may co-reside on the same physical hardware. This deployment setting has motivated a large body of bit-flip attacks (BFAs) against LLMs (including DNNs)~\cite{liu2017fault,breier2018practical,hong2019terminal,yao2020deephammer,rakin2020tbt,tol2023don,chen2021proflip,li2024yes,li2025rowhammer,li2025backdoor,dasattentionbreaker,xusilentstriker,khalil2025flipllm,guo2025sbfa,almalky2025ghosting,lin2025gpuhammer}. Among them, practical BFAs leverage Rowhammer~\cite{kim2014flipping} to demonstrate their end-to-end attacks.
However, existing BFAs have largely focused on corrupting the \emph{model computation}. Most target model parameters or weights and thus require knowledge of the victim model's architecture, weights, or quantization representation to identify fault-vulnerable bits~\cite{yao2020deephammer,rakin2020tbt,chen2021proflip,dasattentionbreaker,xusilentstriker,khalil2025flipllm,guo2025sbfa}.

A smaller body of work instead targets executable code~\cite{li2024yes,chen2025compiledmodels}. Still, they operate on the \emph{computation plane} of DNN inference. Particularly, FrameFlip~\cite{li2024yes} faults a precompiled numerical library (i.e., OpenBLAS). Compiled Models~\cite{chen2025compiledmodels} faults the \texttt{.text} section of model executables generated by DL compilers. While they target different code surfaces, both corrupt code that directly implements model computation. Thus, their demonstrated effects are limited to \emph{inference depletion}, where faulty computation results in corrupted model outputs, making such attacks detectable by existing output-based defenses~\cite{tahmasivand2025lm,zheng2025save,chen2025bitshield}.

JIT-compiled LLM serving introduces a fundamentally different code surface where a \emph{CPU host-side serving control plane} that orchestrates compiled computation rather than implementing the model arithmetic itself. This control logic determines \emph{which compiled artifact serves the current execution} and \emph{whether and how the selected computation is issued to the GPU}. In the PyTorch compiler stack, the control plane spans across different components (e.g., Dynamo frame dispatch).
Thus, while prior executable-code BFAs corrupt the code that \emph{performs} model computation, JIT-compiled serving exposes control logic that {orchestrates} its execution.

This motivates our work which shifts the BFA target from \emph{model computation} to \emph{serving control plane}. We observe that faulting the control plane can disrupt GPU inference while keeping model weights and GPU kernels intact. More importantly, it can keep the model output correct as the faulted execution is redirected to a semantically valid but significantly more expensive path, enabling a \emph{correct-output sponge attack}~\cite{shumailov2021sponge}. Unlike inference depletion, this attack preserves model semantics and is thus difficult to detect from output correctness alone.

To systematically study this unexplored attack surface, we present \name, the first BFAs against the JIT serving control plane of LLM inference. We focus on two families of inference-critical decisions implemented by host-side JIT control code. \emph{Compiled-artifact selection} determines which compiled execution path serves the current inference. \emph{GPU-work submission} determines whether and how the selected computation is issued to the GPU.
Consequently, faulting the former logic can redirect inference onto a much slower but semantically valid execution path, producing the correct-output sponge effect. Faulting the latter can suppress or misdirect required GPU computation and produce \emph{gibberish outputs} while keeping the process alive.

The key challenge is to identify exploitable fault targets in a large and complex JIT compiler stack, where a substantial amount of code implements compilation bookkeeping, error handling, and other auxiliary control rather than inference-critical decisions. FrameFlip~\cite{li2024yes} searches a fault target by exhaustively enumerating and testing every conditional branches within a specific computating function (i.e., \texttt{cblas\_dgemm}). Thus, this search approach does not scale to JIT-compiled LLM serving where the control plane spans multiple compiler and runtime components. Also, an exploitable target should not only alter serving behavior, but also preserve continued execution rather than crashing the execution. It is prohibitively inefficient to exhaustively faulting and evaluating this much larger search space.

To address this challenge, \name employs a \emph{decision-guided fault-vulnerable code analysis} against the targeted PyTorch compiler stack. We first manually reconstruct the JIT serving lifecycle and localize control logic associated with \emph{compiled-artifact selection} and \emph{GPU-work submission}. This semantic characterization reduces the analysis scope from 181 files and 139.1K SLoC in the host-side JIT stack to 34 files and 34.7K SLoC of decision-relevant control logic. We further abstract recurring conditional-control patterns that contribute to the decision-making logic. Guided by these patterns, we propose an automated LLVM Intermediate Representation (IR) analysis to examine the reduced files, resulting in 36,861 conditional branches.
The analysis then builds separate ranking lists for the two attack objectives by assessing how strongly each resulted branch is associated with inference-critical control decisions that can lead to \emph{gibberish output generation} or \emph{sponge effect}. For top-20 from each list, \name identifies 5 output-corruption branches and 10 latency-amplification branches.

We evaluate the severity and transferability of the identified branch targets across four LLMs: Qwen3-8B and DeepSeek-R1-Distill-Qwen-7B on text-generation workloads based on WikiText-2, and Gemma-3-4B and Qwen2.5-VL-7B on multimodal workloads based on ScienceQA. The identified targets produce {gibberish outputs} with PPL ratios ranging from $15.45\times$ to $2.48{\times}10^{6}\times$, or demonstrate a sponge attack by increasing the request latency by $2.03\times$ to $181.90\times$. The branch faults also exhibit cross-model transferability: multiple branches remain effective across different LLMs and workloads while preserving both attack effects.
Further, we evaluate \name against two complimentary BFA defenses for LLMs, FaR~\cite{nazari2024forget} and LM-Fix~\cite{tahmasivand2025lm}. \name successfully bypasses both defenses and retains both attack effects. These results expose a protection gap in existing BFA defenses for LLMs, as they do not cover the JIT serving control plane targeted by \name.

Last, we demonstrate the exploitability of the identified branch targets through end-to-end Rowhammer attacks. Importantly, \name induces bit flips entirely in CPU-resident branch target, and the resulting corruption of JIT serving decisions propagates across the CPU--GPU boundary and disrupts GPU-executed LLM inference, without requiring direct access to GPU memory.
Across all four LLMs, the Rowhammer attacks produce gibberish outputs with PPL ratios from \(287.3\times\) to \(7.23\times10^{6}\), or demonstrate a \emph{correct-output sponge attack} that preserves the generated output while increasing the request latency by \(97.38\times\) to \(124.97\times\).

\section{Background and Related Work}

\subsection{JIT-Compiled LLM Serving}
\label{sec:bg_jit_compiled_inference}
LLM inference consists of {prefill} and {autoregressive decoding}. Prefill processes the prompt and builds the key-value cache. Decoding repeatedly generates new tokens using the cached states. In online serving, requests arrive and finish dynamically, so runtime conditions such as batch size and tensor shapes may vary across executions even for the same model.

Modern LLM serving is predominantly GPU accelerated. The GPU performs the model computation, while the host CPU selects and prepares the computation and submits GPU work. CUDA Graph can further reduce repeated launch overhead by capturing a sequence of GPU operations and replaying it as a reusable graph.

JIT compilation adds compilation and reuse to this execution path. A computation graph can be compiled into a cached \emph{compiled artifact} under runtime conditions recorded as \emph{guards}. On subsequent executions, the runtime evaluates these guards to determine whether an existing artifact can be reused, otherwise, it may select another artifact, recompile, or fall back to a more general path.

\subsection{Rowhammer}
In our attack, \name uses Rowhammer as the primitive to induce a single-bit fault in LLM inference runtime. Rowhammer is a software-induced hardware fault in Dynamic Random-Access Memory (DRAM) that can corrupt memory without directly writing. We briefly review DRAM organization and refer readers to prior work for more details~\cite{kim2014flipping,zhang2022retrospective}.

A DRAM chip is partitioned into banks that organize memory cells as rows and columns. Each cell stores one bit as electrical charge; because the charge leaks over time, DRAM cells require periodic refresh, e.g., every 64\,ms in DDR3 and DDR4.

Kim \textit{et al.}~\cite{kim2014flipping} showed that an unprivileged process can accelerate charge leakage in physically nearby rows and induce bit flips by repeatedly accessing DRAM rows. The frequently activated row is the aggressor row, while the row where bit flips occur is the victim row. A hammering pattern specifies how aggressor rows are selected and accessed; for example, double-sided hammering alternates between two aggressor rows neighboring a victim row to maximize disturbance before the next refresh restores the victim cells' charge.

\subsection{BFAs against DNN/LLM}
\label{sec:related_BFAattack}
BFAs have been comprehensively studied against deep neural networks (DNNs), particularly discriminative models for classification tasks~\cite{liu2017fault,breier2018practical,hong2019terminal,yao2020deephammer,rakin2020tbt,tol2023don,chen2021proflip,li2024yes,li2025rowhammer,li2025backdoor}. Recently, the growing deployment of LLMs has motivated a new line of BFAs targeting generative-model inference. We organize existing BFAs by the objects they corrupt: \emph{model parameters/weights} or \emph{executable code}, with the latter most related to \name.

\mypara{Flipping model parameters/weights}
This line of prior work requires model-specific knowledge (e.g., model weights and architecture) to identify vulnerable parameter or weight bits whose corruption can induce different adverse effects on model behaviors. The most common objective is \emph{inference depletion}, where bit faults degrade inference accuracy or generation quality~\cite{dasattentionbreaker,xusilentstriker,khalil2025flipllm,guo2025sbfa,almalky2025ghosting,lin2025gpuhammer}. A few works induce more selective semantic failures, including safety-alignment bypass and harmful generation~\cite{coalson2024prisonbreak,yan2026has}, attacker-specified outputs~\cite{guo2026tfl,wang2025scalefactors,ahmed2024deeptroj}, and backdoors~\cite{li2025backdoor}; other BFAs can also increase inference cost, e.g., by perturbing EOS-related model state to delay generation termination~\cite{yan2025bithydra}.

Some works only propose algorithms for identifying and emulating vulnerable bit flips without considering hardware and system constraints and therefore do not implement end-to-end attacks~\cite{dasattentionbreaker,xusilentstriker,khalil2025flipllm,guo2025sbfa,almalky2025ghosting,yan2025bithydra}. In contrast, other works~\cite{hong2019terminal,yao2020deephammer,tol2023don,li2025rowhammer,coalson2024prisonbreak} demonstrate end-to-end BFAs, mostly against models residing in CPU DRAM, while GPUHammer~\cite{lin2025gpuhammer} induces bit flips in discrete GPU GDDR.

\mypara{Flipping executable code}
BFAs targeting executable code remain relatively scarce.
More importantly, existing code-targeting attacks and \name differ not only in \emph{where} their target code originates, but also in \emph{what role} that code plays during inference. Existing works target executable code in the \emph{model-computation plane}, whereas \name targets the \emph{serving-control plane} introduced by JIT compilation.

FrameFlip~\cite{li2024yes} identifies vulnerable control-flow instructions in precompiled numerical libraries, e.g., \texttt{cblas\_dgemm} in OpenBLAS, making the proposed BFA model-independent and transferrable across different models.
Instead, Compiled Models~\cite{chen2025compiledmodels} targets the \texttt{.text} section of DNN executables generated by Deep-learning compilers such as TVM and Glow, where model computation is materialized into model-structure-specific native code. Both works demonstrate end-to-end Rowhammer attacks that corrupt executable code in CPU DRAM and deplete CPU-side model inference. In both cases, the fault propagates by changing the numerical computation that produces the model output, restricting their faulting effect to inference depletion.

Unlike them, \name uncovers \emph{JIT-generated serving-control code} as a new attack surface in modern compiled LLM serving. Besides inducing gibberish outputs, \name can also trigger a stealthier \emph{correct-output sponge attack}, significantly increasing inference latency while preserving the model output. More importantly, its CPU-side faults are propagated to affect GPU-executed inference, better reflecting LLM serving environments where GPU inference is predominant.

\subsection{BFA Defenses for DNN/LLM}
\label{sec:related_BFAdefense}
Generally, existing BFA defenses can be categorized by the inference state they protect. Most prior work~\cite{nazari2024forget,nazari2025faraccel,wang2023aegis,chen2025bit} focus on \emph{model state}, such as weights, parameters, activations, and intermediate tensors, using robustness-oriented training, redundant copies for integrity checking and recovery, runtime range or consistency checks, or stronger protection for fault-sensitive values. These defenses are designed to tolerate, detect, or repair corrupted model data. Thus, they mainly address the aforementioned parameter/weight-targeting BFAs, and do not cover faults in the executable control flow that determines how inference is carried out.

A second line of defenses~\cite{tahmasivand2025lm,zheng2025save} detects corrupted inference through \emph{model semantics or output integrity}. These approaches are effective against inference-depletion BFAs that corrupt a model's
prediction or internal decision process. However, this detection signal is insufficient for the sponge attack exposed by \name where the model continues to produce the correct output, but the execution is silently redirected to a significantly more expensive path. Detecting this attack requires monitoring serving-time execution behavior.

BitShield~\cite{chen2025bitshield} provides protection against both weight- and code-targeting BFAs within compiler-generated DNN executables. Its design instruments the DNN compilation pipeline and protects the resulting executable and its semantics. Consequently, its current design does not cover either precompiled shared-library targeted by FrameFlip~\cite{li2024yes} or the JIT serving-control logic targeted by \name, although its principles of semantic- and code-integrity protection could potentially be extended with substantial redesign to protect code-targeted BFAs. More importantly, the redesign would need to consider \name's sponge attack which does not change model semantics.

\section{Threat Model}
\label{threat_model}
Our threat model is consistent with prior Rowhammer
attacks against machine-learning-as-a-service
~\cite{hong2019terminal,yao2020deephammer,li2024yes}. We assume a
software-only, unprivileged attacker that co-resides on the same
physical machine as the victim LLM inference service and uses
Rowhammer to corrupt the integrity of memory-resident code. Thus the underlying DRAM is assumed to be susceptible to Rowhammer-induced bit flips. Such
co-residence can arise in shared servers, multi-tenant cloud
infrastructure, and ML-as-a-service
deployments. The OS and hypervisor are assumed to be correct and enforce process isolation and memory protection. Therefore, the attacker cannot read or write the victim's address space, cannot modify the victim's model files, and cannot directly control the victim's inference requests.

The victim runs a GPU-based LLM inference service with JIT compilation enabled. In our attack, we target the PyTorch compiler stack. We also assume that the victim application, model, and PyTorch software stack contain no software vulnerabilities.
The attacker requires no knowledge of the victim model including its architecture, weights, training data, etc. The attacker relies on knowledge of the PyTorch JIT compiler stack used by the victim, whose source code are publicly available and can be analyzed offline~\cite{pytorchRepository291,anselPyTorch2Faster2024,vllmTorchCompile2025,vllmCompilationConfig2026}. The attacker identifies vulnerable control-flow targets in the PyTorch serving stack offline and later induces faults in the corresponding CPU-resident JIT control code.

\section{\name}
\subsection{Overview}
\label{sec:method_overview}
Our goal is to alter the behavior of JIT-compiled LLM inference with a single-bit fault, without knowing the victim model's architecture, weights, training data, or input prompts. We target two attack effects: \emph{gibberish-output generation}, which corrupts the model output, and \emph{correct-output sponge}, which preserves the output while significantly increasing inference latency~\cite{shumailov2021sponge}.

JIT compilation is particularly relevant to LLM serving because autoregressive generation repeatedly exercises the model execution path~\cite{yu2022orca}. Recurring host-side work is repeatedly exercised along this path, making compiled execution attractive for reducing such overhead while retaining runtime adaptation to changing execution conditions. This execution mode has been integrated into widely used LLM serving software: for example, vLLM V1 uses \texttt{torch.compile} for inference and combines compiled-code caching, Inductor-based compilation, shape specialization, and CUDA-Graph execution~\cite{vllmTorchCompile2025,vllmCompilationConfig2026,vllmCudaGraphs2026}.

In our work, we target the PyTorch compiler stack~\cite{pytorchRepository291} as the JIT implementation. Its \texttt{torch.compile} pipeline provides a representative host-controlled lifecycle spanning TorchDynamo, TorchInductor, Triton-generated GPU kernels, generated wrappers and launchers, and CUDA-Graph replay~\cite{anselPyTorch2Faster2024}. We find that the host-side control code implementing this compiled execution lifecycle also creates a fault-attack surface: flipping a single bit can alter an inference-critical decision and induce either attack effect.
Demonstrating either effect requires connecting a logical code target to a physically flippable memory location. However, the JIT stack contains many control-flow operations, most of which implement auxiliary checks, or error handling rather than inference-critical control. Exhaustively identifying the fault-vulnerable code logic via manual inspection would require substantial effort.

To this end, our decision-guided analysis combines manual semantic characterization with automated static analysis. We first manually localize the host-side control logic from the PyTorch compiler stack that is responsible for compiled-artifact selection and GPU-work submission. We then abstract five recurring conditional-branch archetypes, each characterized by predicate provenance, successor-region behavior, and JIT control context. Guided by these archetypes, we automatically search branches using LLVM IR~\cite{LlvmLlvmproject2026}. The search reconstructs predicate provenance and successor-region behavior, combines them with the localized JIT context to filter and rank candidate branches, and maps the top-ranked branches to binary-level flip records containing a candidate branch's code page, in-page bit offset, and required flip direction.

On the hardware side, Rowhammer-induced bit flips are sparse and depend on the DRAM chips and hammering conditions. Thus, we first profile physical memory to identify flippable bits and record their physical pages, in-page offsets, and flip directions (e.g., 0→1). These profiling records are matched against the binary-level flip records produced above. A candidate branch is exploitable only when a profiled bit has the same in-page offset and flip direction required by the branch record. Once a match is found, we place the target code page onto the matched physical page and trigger the profiled Rowhammer fault to flip the targeted branch bit. This branch inversion changes the corresponding execution-critical decision and produces either gibberish outputs or the correct-output sponge effect, depending on the targeted branch.

\subsection{Execution-Critical Decisions in JIT LLM Serving}
\label{sec:jit_decisions}
From Section~\ref{sec:bg_jit_compiled_inference}, JIT-compiled LLM serving has two properties. First, compiled artifacts are repeatedly reused across model executions, while runtime conditions can change across requests and decoding iterations. So the host runtime must decide whether and which cached artifact should serve each execution. Second, while the model computation itself executes mainly on the GPU, the host CPU is responsible for selecting the computation and submitting it to the GPU.

This forms a host-controlled CPU--GPU execution pipeline. On the CPU, the JIT runtime dispatches model execution, evaluates guards, selects cached compiled artifacts, and prepares their execution. The selected computation is then issued to the GPU through kernel launches or CUDA-Graph replays, where the actual computation is performed. Results produced by the GPU are subsequently consumed by the serving runtime and, during decoding, feed the next generation iteration.

Consequently, this CPU--GPU collaboration exposes the key observation behind \name: while computationally intensive model operators execute on the GPU, {which computation executes and how it is issued
remain governed by host-side control code}. We identify two families of execution-critical decisions as follows.

\mypara{Compiled-artifact selection}
The first decision determines which compiled artifact, if any, should serve the current execution. The host runtime must determine how the execution is dispatched, whether the guard conditions associated with a cached artifact remain satisfied, and whether a matching artifact exists in the compiled cache. If a valid artifact is found, its optimized code can be reused, otherwise, the runtime may try another cached artifact, trigger recompilation, or fall back to a more general execution path. In the PyTorch compiler stack, \texttt{torch.compile} relies on TorchDynamo to intercept model execution, while Dynamo frame dispatch, guard evaluation, and compiled-cache lookup implement the decision regarding whether and which compiled artifact is selected for reuse.

\mypara{GPU-work submission}
Selecting a compiled artifact determines {what} computation should execute, but does not itself execute that computation. The host CPU must still prepare and issue the corresponding work to the GPU. In the PyTorch compiler stack, TorchInductor lowers the selected computation to GPU-side implementations, while generated wrappers and launchers prepare kernel arguments and launch state and submit the resulting Triton kernels. When CUDA Graphs are enabled, host-side runtime logic additionally selects and initiates replay of captured GPU graphs. These decisions determine {whether and how} the selected computation is executed on the GPU.

\subsection{Decision-Guided Fault-Vulnerable Code Analysis}
\label{sec:vulnerable_code_analysis}

The two decision families identified above define the source-level scope of our analysis. JIT compilation introduces a large body of host-side code, much of which performs compilation bookkeeping, auxiliary checks, error handling, or other tasks unrelated to the execution of an inference request. Exhaustively analyzing this entire stack would introduce substantial irrelevant code. Considering only a fraction of the stack can influence the two execution-critical decisions, we use these decision families as semantic anchors to progressively narrow the analysis scope from the full host-side JIT stack to the fault-vulnerable control logic with direct control over them. So we first localize the host-side source control logic responsible for compiled-artifact selection and GPU-work submission below.

\subsubsection{Decision-Relevant Source Localization}
\label{sec:jit_code_targets}
Specifically, we localize the corresponding implementations through a differential
source-analysis procedure below on the PyTorch framework (version \texttt{2.9.1}).

\mypara{Stage responsibility characterization}
We use execution without JIT compilation only as a baseline for isolating JIT-dependent host-side control. We refer to this mode as \emph{eager execution}, where the same inference request follows the framework's regular execution path without relying on JIT-compilation-specific artifacts or runtime state. Our target deployment setting remains JIT-compiled inference.

Starting from the model-execution entry point and ending at GPU-work submission, we inspect the framework source for the same inference path with and without JIT compilation. We reconstruct the host-side control responsibilities in each mode, denoting a lifecycle stage by a distinct control responsibility rather than by every function appearing on the call path. We retain a stage as \emph{JIT-dependent} if its control responsibility is absent from eager execution and its compiled-mode behavior depends on JIT-specific artifacts or runtime
state. The resulting set of stages is denoted by $\mathcal{L}_{\mathsf{dep}}$.

For every stage $s\in\mathcal{L}_{\mathsf{dep}}$, we record four pieces of source-level evidence: (1) its source anchor and entry routine, (2) the JIT-specific artifact or runtime state consumed by the stage, (3) the control result it produces, and (4) how that result affects subsequent request execution. Based on this evidence, we assign each stage a control responsibility $\rho(s)$.

\mypara{Decision-relevance filtering}
We next determine which reconstructed stage responsibilities influence one of the two target decisions. Let
\(\mathcal{D}=\{d_{\mathsf{sel}},d_{\mathsf{sub}}\}\) denote
\emph{compiled-artifact selection} and \emph{GPU-work submission},
respectively. A JIT-dependent stage is relevant to
\(d_{\mathsf{sel}}\) if its control result determines how execution is
dispatched, whether a compiled artifact remains valid, which cached
artifact is selected, or whether execution proceeds through reuse,
compilation, or fallback. A stage is relevant to \(d_{\mathsf{sub}}\)
if its control result determines how the selected computation is
prepared, launched, or replayed on the GPU.
We express these criteria using the predicate \(\operatorname{Relevant}(\rho(s),d)\), which evaluates whether the reconstructed control responsibility \(\rho(s)\) of stage \(s\) contributes to decision \(d\). The predicate is evaluated from the source-level evidence established during the stage responsibility characterization. For each stage that is JIT-dependent, we define \(s\in\mathcal{L}_{\mathsf{dep}}\), and thus
\(
D_s =
\left\{
d\in\mathcal{D}
\mid
\operatorname{Relevant}(\rho(s),d)
\right\}.
\)
We retain a stage if it contributes to at least one of the two target
decisions, resulting in
\(
\mathcal{L}_{\mathcal D}
=
\left\{
s\in\mathcal{L}_{\mathsf{dep}}
\mid
D_s\neq\emptyset
\right\}.
\)
A stage whose output only records, propagates, or maintains auxiliary
state, without determining either target decision, is excluded, i.e.,
\(D_s=\emptyset\).

Algorithm~\ref{alg:decision_source_localization} summarizes this filtering and source-resolution procedure. For every retained stage, \(\operatorname{ResolveImpl}(s,D_s,\mathcal{S})\), where \(\mathcal{S}\) is the framework source tree, resolves its stage-level source anchor to the concrete routines implementing its decision-relevant control behavior. Starting from the source anchor and entry routine identified during the characterization of stage \(s\), it follows framework interfaces, call relationships, and the propagation of JIT-specific runtime state toward the decisions in \(D_s\). A routine is retained if its control behavior directly contributes to at least one such decision, e.g., by evaluating a decision condition or selecting among alternative execution paths; routines that only forward values or maintain auxiliary state, such as logging and bookkeeping routines, are excluded. If the decision-relevant behavior is distributed across multiple routines, each contributing routine is retained. Thus, \(\operatorname{ResolveImpl}(s,D_s,\mathcal{S})\) returns the source-file/routine pairs ($f$, $r$) implementing the decision-relevant control behavior of stage $s$.

The resulting decision-relevant source scope is defined as
\[
\mathcal{F}
=
\left\{
\langle f,r,s,D_s\rangle
\;\middle|\;
s\in\mathcal{L}_{\mathcal D},
\;
(f,r)\in
\operatorname{ResolveImpl}(s,D_s,\mathcal{S})
\right\},
\]
where \(f\) and \(r\) denote the source file and concrete routine,
respectively. Every routine in \(\mathcal{F}\) has a
source-level control role in at least one target decision associated
with its stage.

\begin{algorithm}[t]
\centering
\footnotesize
\begin{algorithmic}[1]
\REQUIRE JIT-dependent stages $\mathcal{L}_{\mathsf{dep}}$;
stage responsibilities $\rho$;
framework source tree $\mathcal{S}$;
decision families
$\mathcal{D}=\{d_{\mathsf{sel}},d_{\mathsf{sub}}\}$

\ENSURE Decision-relevant stages $\mathcal{L}_{\mathcal D}$
and source scope $\mathcal{F}$

\STATE $\mathcal{L}_{\mathcal D} \gets \emptyset$; $\mathcal{F} \gets \emptyset$

\FOR{each stage $s \in \mathcal{L}_{\mathsf{dep}}$}

    \STATE $D_s \gets
    \{d \in \mathcal{D}
    \mid
    \operatorname{Relevant}(\rho(s),d)\}$

    \IF{$D_s \neq \emptyset$}

        \STATE $\mathcal{L}_{\mathcal D}
        \gets
        \mathcal{L}_{\mathcal D} \cup \{s\}$

        \FOR{each $(f,r) \in
        \operatorname{ResolveImpl}(s,D_s,\mathcal{S})$}

            \STATE $\mathcal{F}
            \gets
            \mathcal{F}
            \cup
            \{\langle f,r,s,D_s\rangle\}$

        \ENDFOR

    \ENDIF
\ENDFOR

\STATE \textbf{return}
$\mathcal{L}_{\mathcal D}, \mathcal{F}$

\end{algorithmic}
\caption{Decision-relevance filtering.}
\label{alg:decision_source_localization}
\end{algorithm}

\mypara{Localized JIT-dependent control stages}
Applying the above procedure reduces the source-analysis scope from {181 files} and {139.1K SLoC} in the host-side JIT stack to {34 files} and {34.7K SLoC} of decision-relevant control logic. For \emph{compiled-artifact selection}, we retain \emph{frame dispatch}, \emph{guard evaluation}, and \emph{compiled-cache lookup}, which determine, respectively, how an intercepted computation is handled, whether the runtime conditions of a cached artifact still hold, and whether a matching artifact is reused or execution proceeds through another artifact, compilation, or fallback.

For \emph{GPU-work submission}, we retain the \emph{generated wrapper and launcher} path, which prepares the state and arguments required to issue the selected computation. This path includes the \emph{Inductor static launcher}, which prepares and issues kernel launches, and the \emph{Triton launcher}, which invokes the GPU kernels. When CUDA Graphs are used, we retain \emph{host-side graph replay}, which determines whether and how a captured sequence of GPU work is replayed. These stage semantics are reconstructed from source-level control responsibilities rather than defined by the framework itself.

\subsubsection{Decision-Relevant Branch Archetypes}
\label{sec:branch_archetypes}
The source localization identifies the JIT-dependent stages and specific routines that participate in the two decisions. However, the resulting source scope \(\mathcal{F}\) contains many control-flow operations, most of which implement local bookkeeping, auxiliary checks, or error handling rather than the decision semantics of interest. Thus, we analyze the decision-relevant source scope \(\mathcal{F}\) at the level of conditional control flow.

Both execution-critical decisions are ultimately made through control-flow choices in the host runtime. For \emph{compiled-artifact selection}, conditional branches determine, e.g., whether an execution follows a compiled or fallback path, whether a guard result is accepted, and whether a cached artifact is selected for reuse. For GPU-work submission, conditional branches determine, e.g., whether a kernel launch proceeds or whether a particular launch path is taken. Thus, conditional branches can provide the control flow through which the decisions are made.
This observation makes a conditional branch a natural flipping target. A branch can select one of two successor paths depending on a runtime predicate. Thus, inverting its branch can  redirect an otherwise unchanged execution to the alternative path and alter the decision implemented by that branch.
Also, it is compatible with single bit-flip. For example, in the x86 conditional-jump encoding, \texttt{jle} (\texttt{0x7e}) and \texttt{jg} (\texttt{0x7f}) differ by
one bit while reversing the branch condition. A single-bit fault at such an opcode position can transform a valid conditional jump into another valid conditional jump, redirecting control flow without introducing an invalid instruction. This is important, as our goal is not to corrupt arbitrary branches, but to identify branches for which the alternative path changes an inference-relevant decision while still preserving continued execution without any crash.

\mypara{From semantic anchors to branch archetypes}
However, exhaustively interpreting the semantics of every branch in \(\mathcal{F}\) would be time-expensive. Instead, we use the reconstructed semantics of the localized stages above to identify a small number of \emph{semantic anchor branches}. Recall that each retained stage is associated with a control responsibility $\rho(s)$, the JIT-specific state it consumes, the control result it produces, and the effect of that result on subsequent execution. Thus, we start from the source anchor of each stage, we follow this stage-specific evidence to the control points at which its responsibility is implemented as a conditional choice.

The identified anchor branches are concrete instances tied to particular source locations. Our objective, however, is to find as many branches as possible that implement the same decision semantics elsewhere in the localized code. As such, we abstract each anchor according to three properties: (1) the provenance of the runtime value tested by its predicate, (2) the behavioral difference between its two successor regions, and (3) the JIT control context in which the branch occurs. Branches with the same combination of the three properties are grouped into a \emph{decision-relevant branch archetype}.
This abstraction generates five recurring archetypes, each capturing a recurring conditional-control pattern through which one of the two decisions is made.

\begin{itemize}[noitemsep, topsep=2pt, partopsep=0pt,leftmargin=0.4cm]
\item \textbf{Compiled-reuse gating.}
\emph{Predicate provenance:} the tested value represents whether an available
compiled execution path can serve the current execution.
\emph{Successor behavior:} the two successor regions distinguish compiled
reuse from additional host-side processing, such as a callback, recompilation, or fallback.
\emph{JIT context:} the branch occurs within frame dispatch or related
transitions into compilation or code generation.

\item \textbf{Guard-result gating.}
\emph{Predicate provenance:} the tested value is derived from guard-evaluation
state and represents whether the runtime conditions required by a compiled
artifact are accepted.
\emph{Successor behavior:} the two successor regions distinguish acceptance
of the current artifact from further guard processing or rejection.
\emph{JIT context:} the branch occurs during guard evaluation.

\item \textbf{Cache-match gating.}
\emph{Predicate provenance:} the tested value is derived from compiled-cache
lookup or match state and represents whether a candidate cache entry matches
the current execution.
\emph{Successor behavior:} the two successor regions distinguish selection
of the matching artifact from continued lookup, a cache miss, recompilation,
or fallback processing.
\emph{JIT context:} the branch occurs during compiled-cache lookup.

\item \textbf{Launch-grid gating.}
\emph{Predicate provenance:} the tested value is derived from a computed
launch configuration, such as the dimensions or derived size of a launch grid.
\emph{Successor behavior:} one successor proceeds toward kernel submission,
whereas the other bypasses or returns without issuing that work.
\emph{JIT context:} the branch occurs within generated wrappers or launchers.

\item \textbf{Launch-path resolution.}
\emph{Predicate provenance:} the tested value represents launch-resolution
state, such as whether a lazy launch handle has been resolved.
\emph{Successor behavior:} the two successor regions distinguish the normal
resolved launch path from a resolution or alternative path.
\emph{JIT context:} the branch occurs within the generated or runtime launcher path.

\end{itemize}

These archetypes seed the automated search as structural signatures over predicate provenance, successor-region behavior, and JIT component context, allowing other branches with the same decision pattern to be identified in the localized source scope without manual semantic inspection of each branch. We denote the conditional branches enumerated from $\mathcal{F}$ by $\mathcal{B}$.

\begin{algorithm}[t]
\footnotesize
\caption{Archetype-guided fault-vulnerable code search.}
\label{alg:search_procedure}
\begin{algorithmic}[1]
\REQUIRE Candidate branches $\mathcal{B}$ from localized source scope
$\mathcal{F}$; archetype signatures $\mathcal{A}$; selection budget $N$
\ENSURE Top-$N$ gibberish and sponge binary flip records

\STATE $\mathcal{R} \gets \emptyset$

\FOR{each candidate branch $b \in \mathcal{B}$}
    \STATE $G_b \gets \mathrm{RecoverLocalCFG}(b)$
    \STATE $S_b \gets \mathrm{BackwardPredicateProvenance}(b,G_b)$
    \STATE $E_b^T,E_b^F \gets \mathrm{ForwardSuccessorBehaviors}(b,G_b)$
    \STATE $J_b \gets \mathrm{JITContext}(b)$
    \STATE $C_b \gets \langle S_b,E_b^T,E_b^F,J_b\rangle$

    \FOR{each archetype $a \in \mathcal{A}$}
        \IF{$\mathrm{CandidateFiltering}(C_b,a)=\mathrm{retained}$}
            \STATE $\mathcal{R} \gets
            \mathcal{R}\cup\{\langle b,a\rangle\}$
        \ENDIF
    \ENDFOR
\ENDFOR

\STATE $\mathcal{R}_{\mathsf{gib}},
       \mathcal{R}_{\mathsf{spg}}
       \gets \mathrm{CandidateRanking}(\mathcal{R})$

\STATE $\widehat{\mathcal{R}}_{\mathsf{gib}}
       \gets \mathrm{TopN}(\mathcal{R}_{\mathsf{gib}},N)$
\STATE $\widehat{\mathcal{R}}_{\mathsf{spg}}
       \gets \mathrm{TopN}(\mathcal{R}_{\mathsf{spg}},N)$

\FOR{each $\langle b,a\rangle$ in
$\widehat{\mathcal{R}}_{\mathsf{gib}}
\cup
\widehat{\mathcal{R}}_{\mathsf{spg}}$}
    \STATE $\ell_b \gets \mathrm{BinaryFlipRecord}(b)$
    \STATE Attach $\ell_b$ to $\langle b,a\rangle$
\ENDFOR

\STATE \textbf{return}
$\widehat{\mathcal{R}}_{\mathsf{gib}}$ and
$\widehat{\mathcal{R}}_{\mathsf{spg}}$
with their $\ell_b$ records

\end{algorithmic}
\end{algorithm}

\subsubsection{Backward Predicate Provenance Analysis}
\label{sec:backward_predicate_analysis}
The first part of the search asks a question: \emph{what runtime value is a given branch actually testing?} A branch condition may appear only as a Boolean value at the control point, but that Boolean can represent the runtime state associated with different archetypes. To distinguish them, we trace the tested value backward to the computations that produced it.

In Lines~3--4 of Algorithm~\ref{alg:search_procedure}, we construct $G_b$ for the routine containing a candidate branch $b$ and apply $\mathrm{BackwardPredicateProvenance}$ to trace the data dependencies of its condition backward within the routine. This process reconstructs the computations that produce the branch condition.

We denote the resulting dependency slice by the \emph{predicate slice} $S_b$, which includes only computations that can affect the branch condition. Along the slice, we retain operations, constants, comparisons, and direct calls that explain where the tested value comes from. For a direct call associated with a retained JIT stage $s$, we annotate it with the stage-level semantics reconstructed during source localization in Section~\ref{sec:jit_code_targets}. Thus, the call provides evidence about whether the predicate depends on, for example, frame-dispatch state, guard evaluation, compiled-cache lookup, or launcher-side control.

The tracing remains within the routine that contains the branch. We do not recursively analyze the bodies of called routines. Instead, values defined outside the routine are treated as boundary inputs, while direct calls associated with localized JIT stages retain the previously reconstructed semantics. This keeps the analysis local while preserving the stage-level evidence needed to interpret the branch predicate.
To bound the per-branch search cost, we impose two limits: a maximum backward depth of $k$ (e.g., 12) and a budget (e.g., 64) of visited computations. Slices truncated by either limit are marked as incomplete, and this information will be used later in Section~\ref{sec:matching}.

The resulting $S_b$ provides the predicate-provenance part of the archetype matching, which additionally considers successor-region behavior and the JIT context in which the branch appears.

\subsubsection{Forward Successor-Region Analysis}
\label{sec:forward_successor_regions}
The forward analysis asks the complementary question: \emph{what can execution do differently after taking the two successors of a branch?} This difference forms the second part of the archetype matching. For a conditional branch $b$, we denote by $E_b^T$ and $E_b^F$ the behavior summaries of its two outgoing successors, corresponding to the predicate evaluating to the two outgoing branch edges. In Line~5 of Algorithm~\ref{alg:search_procedure}, we apply $\mathrm{ForwardSuccessorBehaviors}$ to the two successors of $b$ over $G_b$, producing $E_b^T$ and $E_b^F$.

The analysis operates on the local control-flow graph $G_b$ (CFG) of the routine containing $b$.

Rather than preserving every instruction reachable from a successor, we represent each summary $E_b^T$ or $E_b^F$ as a finite set of \emph{behavior facts} describing behavior that may occur after taking that successor. These facts fall into three categories: (1) completion or termination behavior; (2) JIT-stage semantics observed through direct calls; and (3) repetition evidence indicating that the successor region contains repeated control flow.

Starting from each successor, we perform bounded, path-insensitive forward reachability over $G_b$~\cite{reps1995precise,nielson2005principles,cousot1977abstract}, visiting at most 128 basic blocks on each side. A behavior fact is recorded if at least one CFG path from the successor can reach it. We do not evaluate the runtime conditions that determine whether that path is actually taken and instead record what {may} be reachable from each successor. When multiple CFG paths merge, the behavior facts collected along those paths are combined by set union.

For completion and termination behavior, a terminal \texttt{ret} contributes a normal-completion fact. An \texttt{invoke} contributes exceptional-exit evidence through its unwind successor, while \texttt{unreachable} contributes abnormal-termination evidence. These facts distinguish a successor from which execution can complete normally from one that leads only to failure.

Direct calls contribute JIT-stage behavior facts. As in the backward analysis in Section~\ref{sec:backward_predicate_analysis}, we do not recursively analyze the bodies of called routines. Instead, when a direct call is associated with a retained JIT stage $s$, we reuse the stage-level semantics. A call contributes a behavior fact corresponding to semantics such as frame dispatch, guard evaluation, compiled-cache lookup, etc. This context-insensitive call abstraction~\cite{sharir1981two} allows direct calls to different routines to be compared through the JIT semantics they represent rather than through their concrete function bodies.

We also record when a successor region contains repeated control flow. We first compute the bounded set of CFG blocks reachable from the successor and then inspect the resulting subgraph for a back edge, i.e., an edge indicating a reachable cycle~\cite{kildall1973unified,tarjan1972depth}. A detected cycle is recorded once as a repetition fact rather than being repeatedly unfolded. This captures whether one successor involves repeated host-side processing. The 128-block budget bounds the per-branch analysis cost. If the budget is reached before the reachable region has been fully explored, we mark the behavior summary for that successor as incomplete, and this information is later used in Section~\ref{sec:matching}.

After constructing $E_b^T$ and $E_b^F$, we compare the two sets directly. Facts in their intersection are reachable from both successors and do not distinguish the outcomes, whereas their symmetric difference captures behavior reachable from only one successor and thus behavior that can change when the branch is inverted. For example, one successor may reach compiled reuse while the other reaches fallback processing, or one successor may reach kernel submission while the other returns without issuing the work. The pair $\langle E_b^T,E_b^F\rangle$ therefore provides the successor-region-behavior part of the archetype matching.

\subsubsection{Fault-Vulnerable Code Identification}
\label{sec:matching}
At this point, the search asks a final question: \emph{does the evidence collected for a candidate branch exhibit the same decision pattern as one of the five archetypes?} In Lines~6--7 of Algorithm~\ref{alg:search_procedure}, we apply $\mathrm{JITContext}$ to $b$ and assemble the recovered evidence as
\(
C_b=\langle S_b,E_b^T,E_b^F,J_b\rangle,
\)
where $S_b$ captures predicate provenance, $E_b^T$ and $E_b^F$ summarize successor-region behavior, and $J_b$ records the JIT control context. The first two forms of evidence are obtained from the backward and forward analyses above. To obtain $J_b$ from the source scope $\mathcal{F}$ in Section~\ref{sec:jit_code_targets}, we use the stage $s$ and routine $r$ associated with the record $\langle f,r,s,D_s\rangle\in\mathcal{F}$ for the routine containing $b$. Thus, $C_b$ represents the same three forms of evidence used to define the archetypes in Section~\ref{sec:branch_archetypes}.

\mypara{Archetype signatures}
Each archetype $a$ is represented by a \emph{structural signature} over the
same three forms of evidence: predicate provenance, successor-region behavior, and JIT component context. For a candidate branch $b$, the search compares $C_b$ against the three checks below.

\emph{First}, a \emph{predicate-provenance check} determines whether $S_b$ contains the value-flow evidence required by archetype $a$. We view $S_b$ as the dependence graph formed by the computations retained during the backward analysis and match archetype-specific patterns over its operations, constants, comparisons, and summarized direct calls~\cite{ullmann1976algorithm}. \emph{Second}, a \emph{successor-contrast check} determines whether $E_b^T$ and $E_b^F$ exhibit the behavioral difference defined by archetype $a$. In particular, the required contrast must be present in the behaviors
that distinguish the two successor sets. \emph{Third}, a \emph{JIT-context check} determines whether $J_b$ places the branch in the localized stage and routine context defined for archetype $a$.
\emph{Last}, we apply a \emph{continued-execution check}. A candidate is retained only when the branch-side behavior relevant to the intended inversion is not represented only by exceptional or abnormal termination. This check enforces our requirement that branch inversion alter an execution-critical decision without crash.

\mypara{Candidate branch filtering}
Let $\mathcal{A}$ denote the five decision-relevant branch archetypes defined in Section~\ref{sec:branch_archetypes}, where each specifies requirements over predicate provenance, successor-region behavior, and JIT control context. The search evaluates each candidate branch $b$ against every $a\in\mathcal{A}$ using its recovered evidence $C_b=\langle S_b,E_b^T,E_b^F,J_b\rangle$.

In Lines~9--10 of Algorithm~\ref{alg:search_procedure}, $\mathrm{CandidateFiltering}$ applies the following rules and adds every retained pair $(b,a)$ to the record set $\mathcal{R}$.
We first apply the \emph{continued-execution check} above to every branch--archetype pair $(b,a)$ and discard pairs that fail it.

For the remaining pairs, we evaluate the three archetype-specific checks. A pair $(b,a)$ is retained as a \emph{direct match} if all three checks are satisfied. If the JIT-context check is satisfied but either of the other two checks is not established, we retain the pair at lower priority as a \emph{context-supported match}. This case preserves branches located in a stage and routine already identified during source localization. All remaining pairs are discarded.
Because the backward and forward analysis are bounded, both direct and context-supported matches may contain incomplete analysis results. This completeness information is used only for ranking: among direct matches, those supported by complete backward and forward analysis are ranked ahead of those with incomplete analysis. Context-supported matches are ranked below direct matches.

\mypara{Candidate branch ranking}
The retained pairs are prioritized separately for the two attack objectives: \emph{gibberish-output generation} and the \emph{correct-output sponge}. In Line~14 of
Algorithm~\ref{alg:search_procedure}, $\mathrm{CandidateRanking}$ first separates the retained candidates into two ranked lists, i.e., $\mathcal{R}_{\mathsf{gib}}$ and $\mathcal{R}_{\mathsf{spg}}$. $\mathcal{R}_{\mathsf{gib}}$ contains candidates associated with the
GPU-work-submission archetypes and targets gibberish-output generation, whereas $\mathcal{R}_{\mathsf{spg}}$ has candidates associated with the compiled-artifact-selection archetypes and targets the sponge attack.

Within each list, direct matches are ranked ahead of context-supported matches because they satisfy all three archetype-specific checks. Within the same match class, candidates for which the backward predicate analysis and both forward successor-region analysis complete are ranked ahead of candidates for which one of the analysis are marked as incomplete. Thus, the ranking  prioritizes the strength of the archetype match and then the completeness of the recovered evidence. In Lines~15--16 of Algorithm~\ref{alg:search_procedure}, $\mathrm{TopN}$ selects the first $N$ records (e.g., $N=20$ in our implementation) from each ranked list, producing $\widehat{\mathcal{R}}_{\mathsf{gib}}$ and $\widehat{\mathcal{R}}_{\mathsf{spg}}$ for the binary-level analysis below.

\mypara{Binary-level fault-vulnerable branch records}
For each branch--archetype pair in $\widehat{\mathcal{R}}_{\mathsf{gib}}\cup\widehat{\mathcal{R}}_{\mathsf{spg}}$, we apply $\mathrm{BinaryFlipRecord}$ in Lines~18--19 of Algorithm~\ref{alg:search_procedure} to locate the corresponding conditional branch in the generated binary and attach its record $\ell_b$. The resulting record $\ell_b$ contains the target binary and code page, the in-4KiB-page bit offset of the vulnerable code bit, and the required flip direction. Thus, this step maps a source-level identified branch to the concrete binary bit location whose flip can invert the corresponding control decision.

\subsection{Memory Profiling}
\label{sec:memory_profiling}

The vulnerable branch bits above provide the code-side targets. A practical attack additionally requires flippable bits in physical memory. Rowhammer-flippable bits are sparse and hardware-dependent, so we perform memory profiling to identify and record such bits before aligning them with vulnerable branch bits by in-page offset and flip direction.

We first leverage prior DRAM-address reverse-engineering techniques~\cite{pessl2016drama,wang2020dramdig} to recover the mapping from physical addresses to DRAM banks and rows. We then use hammer-pattern fuzzing techniques~\cite{frigo2020trrespass,jattkeblacksmith} to search for an effective hammering pattern on the target machine, including patterns that bypass Target Row Refresh (TRR) on modern DRAM chips. With the selected pattern, we allocate attacker-owned memory, run Rowhammer tests on candidate physical pages, and record each observed flip, including the flipped physical page, the bit offset within the 4\,KB page, the flip direction, and the aggressor pages that produced the flip.

The target branch bit must have the same in-page offset as a profiled flippable bit, and the observed flip direction must match the direction required to transform the branch opcode into its opposite successor. A profiled bit that does not satisfy both conditions cannot be used for the selected branch inversion.
We note the profiling stage is independent of the victim model and the victim process. It only produces records of flippable locations.

\subsection{Targeted Runtime Bit Flip}
\label{sec:rowhammer_exploitation}

Given a vulnerable branch record and a profiled physical location matched by in-page bit offset and flip direction, the remaining step is to place the branch's file-backed code page onto the matched physical page and trigger the bit flip.

The target branch resides in a file-backed executable page of the JIT compiler stack. Since the attacker cannot overwrite this page or choose its physical frame, we use an established memory waylaying technique~\cite{gruss2017another} to force the OS to reload the target code page onto the matched flippable physical page. After the placement succeeds, the attacker applies the profiled hammering pattern to the corresponding aggressor pages. The resulting bit flip modifies the selected opcode bit, converts the conditional branch into its opposite form, and causes subsequent inference requests to run with the corrupted branch decision.

\section{Evaluation}
\label{sec:evaluation}

\subsection{Experiment Setup}
\label{sec:eval_setup}

\mypara{Hardware platform}
We use two hardware platforms in our evaluation. The first platform is a GPU server running Ubuntu 20.04.6 with two Intel Xeon E5-2680 v4 CPUs, 128 GiB Samsung DDR4 ECC memory, and five NVIDIA GeForce RTX 3090 GPUs, each with 24 GiB of memory. The server uses NVIDIA driver 550.54.14 and CUDA 12.3. We use this platform to assess the fault-vulnerable attack surface. The second platform is a machine running Ubuntu 20.04.6 with an Intel Core i3-10100 CPU and 16 GiB non-ECC Apacer DDR4-2133 memory installed as two 8 GiB DIMMs in a dual-channel configuration. It is equipped with one NVIDIA GeForce RTX 3090 GPU with 24 GiB of memory and uses NVIDIA driver 575 and CUDA 12.3. On this platform, the recovered DRAM address
mapping computes the bank index from XORs over the physical-address
bit pairs \((17,21)\), \((16,20)\), \((15,19)\), \((14,18)\), and
\((6,13)\), and derives the row index from physical-address bits
$18--33$. The effective hammering pattern was a 4-sided hammer pattern. Using this setup, we observed 137 bit-flip events across 49 victim physical pages, including 63 \(1{\rightarrow}0\) flips and 74 \(0{\rightarrow}1\) flips. For each observed flip, we recorded the victim physical page, the bit offset within the 4\,KB page, the flip direction, and the corresponding aggressor pages. We use this platform to perform end-to-end Rowhammer attacks.

\mypara{LLM inference framework}
We implement \name on PyTorch 2.9.1 using the \texttt{torch.compile}--Inductor--Triton stack with CUDA Graph support~\cite{anselPyTorch2Faster2024}. This stack is representative of modern JIT-compiled LLM serving.

\mypara{LLM models and workloads}
We evaluate two text-only models, Qwen3-8B~\cite{QianWen38B} and DeepSeek-R1-Distill-Qwen-7B~\cite{DeepSeekR1DistillQwen7B}, and two multimodal models, Gemma-3-4B~\cite{Gemma34bIt} and Qwen2.5-VL-7B~\cite{QianWen25VL7BInstruct}. We refer to them as Qwen3, DeepSeek-R1, Gemma3, and Qwen2.5-VL. This allows us to evaluate \name's transferability in different model families.

For output-corruption (resulting in gibberish output) measurements on the text-only models, we use the first 512 tokens of WikiText-2~\cite{MindchainWikitext2Datasets2025} as a fixed reference for PPL and retain a fixed 32-token generation as visible output evidence. For the multimodal models, we use three fixed image-based multiple-choice questions from ScienceQA~\cite{ScienceQAScienceQuestion} (pid 591, 570, and 278). Each question has four answer choices, and we compute PPL against the correct answer option. We aggregate the three questions by averaging their correct-choice NLLs before converting to PPL, equivalently taking the geometric mean of their PPL values.

For inference latency measurements, each model uses one fixed generation request. The text-only workload contains a 128-token prompt followed by a fixed 32-token decode. The multimodal workload uses ScienceQA pid 591 with the same fixed 32-token decode. For each attack effect, the four model-specific workloads constitute four \emph{settings}. These fixed workloads provide a stable basis for measuring clean-to-faulted metric changes and for evaluating whether the same branch remains effective across models.

\mypara{Metrics}
We quantify output corruption using the PPL ratio and latency amplification using the Latency ratio:
\[
  \mathrm{PPL\,ratio}
  = \frac{\mathrm{PPL}_{\mathrm{faulted}}}
         {\mathrm{PPL}_{\mathrm{clean}}},
  \qquad
  \mathrm{Latency\,ratio}
  = \frac{T_{\mathrm{faulted}}}
         {T_{\mathrm{clean}}},
\]

where PPL is the perplexity for text-only inference or the perplexity computed over the answer choices for multimodal inference, and $T$ is the wall-clock time of one end-to-end inference request on the fixed workload, measured from request submission to the completed response and averaged over repeated runs. A ratio of $1\times$ means no change. For PPL, we treat a completed run as corrupted when its PPL ratio reaches at least $10\times$; non-finite values are treated as exceeding this threshold. For latency, larger ratios indicate multiplicative slowdowns. Both ratios normalize the clean-to-faulted change within each setting, because raw PPL values and raw request latencies differ across tokenizers, workloads, models, and hardware. Outcomes that do not yield a numeric metric, such as crashes or explicit errors, are tracked separately. For readability, plots cap extremely large numeric ratio values, and branch examples reported later are drawn from non-crashing numeric outcomes.

\subsection{Attack Surface Assessment}
\label{sec:eval_attack_surface_assessment}

\subsubsection{Fault-vulnerable Code Search Efficiency}
\label{sec:eval_branch_analysis}
We evaluate whether \name can efficiently prioritize vulnerable branch bits compared with uniform random sampling and brute-force sequential traversal under the same fault-injection budget.
Our automated LLVM IR analysis identifies 36{,}861 conditional branches in the localized JIT code. \name then ranks these branches separately for gibberish-output generation and latency amplification according to their association with the corresponding execution-critical decisions. For each objective, we test the Top-20 ranked candidates and compare them with uniform random sampling and sequential traversal over the same branch corpus under an identical 20-trial budget. Random sampling is repeated with three independent seeds.

We classify a branch as vulnerable to the gibberish attack if its flip yields a PPL ratio of at least $10\times$ in at least one evaluated setting. A branch is vulnerable to latency amplification if its flip yields a Latency ratio of at least $2\times$ in at least one evaluated setting. As shown in \Cref{tab:branch_budget_compare}, under the same budget, the sequential baseline and each of the three independent random trials discover 0 vulnerable branches for either effect, whereas our algorithm discovers 5 branches capable of inducing gibberish and 10 capable of amplifying latency, yielding precisions of 25.0\% and 50.0\%, respectively. These results show that the semantic ranking concentrates vulnerable branches within the early candidate set.

\begin{table}[!t]
\centering
\caption{Results of confirmed vulnerable targets under a 20-trial budget for different methods.}
\label{tab:branch_budget_compare}
\footnotesize
\setlength{\tabcolsep}{3pt}
\begin{tabular}{lccc}
\hline
Method & Budget & Targets Found & Precision \\
\hline
\multicolumn{4}{l}{\textit{Output corruption}} \\
\hline
Random & 20 & 0 & 0\% \\
Sequential & 20 & 0 & 0\% \\
Ours & 20 & \textbf{5} & \textbf{25.0\%} \\
\hline
\multicolumn{4}{l}{\textit{Latency amplification}} \\
\hline
Random & 20 & 0 & 0\% \\
Sequential & 20 & 0 & 0\% \\
Ours & 20 & \textbf{10} & \textbf{50.0\%} \\
\hline
\end{tabular}
\end{table}

The consistent absence of confirmed targets in random sampling and brute-force sequential traversal suggests that vulnerable branch bits are sparse in the analyzed JIT-compiled code. The higher hit rates of our Top-20 lists are consistent with the search design: it ranks branches according to the runtime decision encoded by the predicate, the consequences of the successor paths, and the surrounding decision context, thereby prioritizing branch inversions likely to propagate into subsequent inference computation.
This efficiency matters because validating a branch through fault-injection trials is costly in both practical attack workflows and framework security auditing. For practical attacks, a smaller candidate set enriched with vulnerable branch bits reduces ineffective trials and makes target discovery feasible under a limited fault-injection budget. For framework developers and maintainers, the same vulnerable-branch search algorithm provides a practical way to prioritize security testing: it highlights branch bits whose predicate decision, successor-path consequence, and decision context connect them to gibberish outputs or latency amplification.

\subsubsection{Attack Severity and Transferability}
\label{sec:eval_transfer}

We evaluate the severity and transferability of the Top-20 branches identified by our search for each attack effect. We flip these branches across the four model--workload settings, measure the resulting attack severity, and count the settings in which each branch remains effective.

\begin{figure}[!t]
\centering
\includegraphics[width=0.8\linewidth]{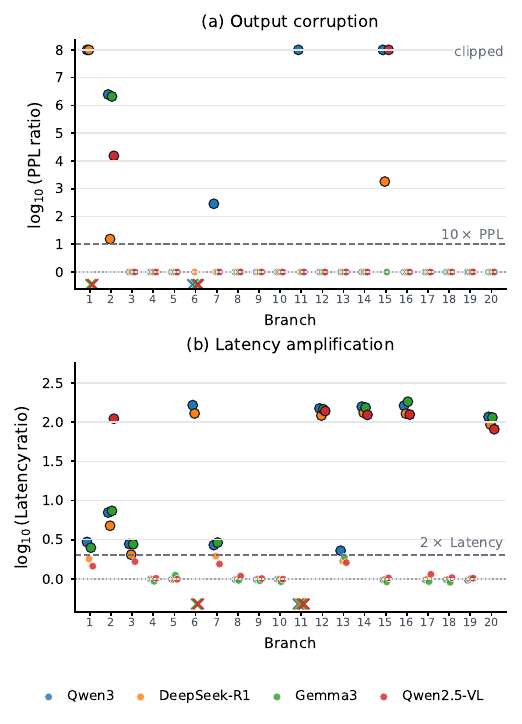}
\caption{Attack severity for gibberish output (top) and latency amplification (bottom). Each point represents one branch--model setting. Branches are ordered by index, and crosses denote abnormal program termination without a completed numeric outcome.}
\label{fig:branch_severity}
\end{figure}

\Cref{fig:branch_severity} separately presents branch-level severity for output corruption and latency amplification. For output corruption, it reports $\log_{10}$ of the PPL ratio; for latency amplification, it reports $\log_{10}$ of the Latency ratio. Each point represents one branch--model setting. On the x-axis, branches are ordered by index. The dotted lines mark the $10\times$ PPL ratio threshold and the $2\times$ Latency ratio threshold used to determine whether a branch achieves the intended attack effect and is considered vulnerable. Values above the output-corruption plotting cap are clipped. A cross indicates that flipping the branch causes the program to terminate abnormally in the corresponding model setting without a completed numeric outcome.

Among the vulnerable-branch outcomes that achieve output corruption, the PPL ratio ranges from $15.45\times$ to $2.48{\times}10^{6}\times$, with an arithmetic mean of $7.65{\times}10^{5}\times$. For latency amplification, the vulnerable-branch outcomes yield Latency ratios from $2.03\times$ to $181.90\times$, with an arithmetic mean of $85.48\times$.

\Cref{fig:branch_severity} shows that neither effect is tied to a single LLM model or workload. Across the analyzed vulnerable-branch candidates, multiple branches produce large PPL ratios or Latency ratios in more than one setting, and several effects exceed the plotting caps. These recurring high-impact outcomes show that vulnerable branches in the code introduced by JIT compilation can remain damaging across different models. Unlike model-parameter faults, our attack targets the code introduced by JIT compilation rather than model parameters. As a result, the attacker does not need to know or modify model parameters, and different models, including multimodal models, may be affected.

\begin{figure}[!ht]
\centering
\includegraphics[width=0.8\linewidth]{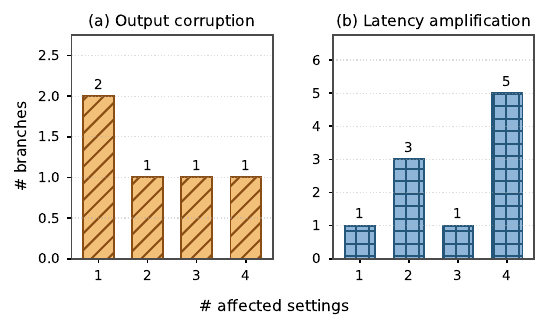}
\caption{Attack transferability for gibberish output (left) and latency amplification (right).}
\label{fig:transferability_summary}
\end{figure}

\Cref{fig:transferability_summary} summarizes branch-level transferability by grouping branches according to the number of settings in which flipping the branch achieves the corresponding attack effect. Subfigure~(a) reports the five vulnerable branches that cause output corruption: 1 affects all four settings, 1 affects three settings, 1 affects two settings, and 2 affect one setting. Thus, 2 out of 5 branches transfer to at least three settings. Subfigure~(b) reports the ten vulnerable branches that amplify latency: 5 affect all four settings, 1 affects three settings, 3 affect two settings, and 1 affects one setting. Thus, 6 out of 10 branches transfer to at least three settings. These results show that both attack effects exhibit transferability across models and workloads.

\subsubsection{Attack Effectiveness against BFA Defenses}
\label{sec:eval_defense}

We next evaluate whether two existing weight-level BFA defenses mitigate our branch faults. FaR~\cite{nazari2024forget} hardens a model offline by rewiring selected linear layers, whereas LM-Fix~\cite{tahmasivand2025lm} detects, localizes, and recovers weight bit flips at runtime. FaR's hardening is expensive: its authors report 237 hours of offline hardening computation for a 334M-parameter model, so hardening our 7B--8B evaluation models would require an estimated hundreds to thousands of hours. We therefore conduct this experiment on the smaller Qwen3-0.6B and use the same model for LM-Fix.

We first verify both defenses against weight-targeting BFAs. The unprotected model reaches a $2\times$ PPL increase after 2 weight-bit flips. After 5 weight-bit flips, the FaR-hardened model remains below this threshold at a $1.97\times$ PPL ratio and crosses the threshold only on the sixth flip. For LM-Fix, all 4 injected weight-bit flips are detected, precisely localized, and recovered, restoring both the original weight value and the golden result. These results show that both defenses resist fault injection into model weights.

\begin{table}[!t]
\centering
\caption{Effectiveness of FaR and LM-Fix against a weight-targeting BFA and \name on Qwen3-0.6B.}
\label{tab:bfa_defenses}
\footnotesize
\setlength{\tabcolsep}{3pt}
\begin{tabular}{@{}l l l@{}}
\hline
Fault & Injection & Outcome \\
\hline
\multicolumn{3}{@{}l}{\textit{Unprotected}} \\
Weight-targeting BFA & 2 weight-bit flips & $2.00\times$ PPL \\
Gibberish branch fault & 1 branch-bit flip & $35{,}831.1\times$ PPL \\
Sponge branch fault & 1 branch-bit flip & $188.0\times$ latency \\
\hline
\multicolumn{3}{@{}l}{\textit{FaR}} \\
Weight-targeting BFA & 5 weight-bit flips & $1.97\times$ PPL \\
Gibberish branch fault & 1 branch-bit flip & $35{,}831.2\times$ PPL \\
Sponge branch fault & 1 branch-bit flip & $206.1\times$ latency \\
\hline
\multicolumn{3}{@{}l}{\textit{LM-Fix}} \\
Weight-targeting BFA & 4 weight-bit flips & 4/4 detected and recovered \\
Gibberish branch fault & 1 branch-bit flip & $37{,}428.9\times$ PPL \\
Sponge branch fault & 1 branch-bit flip & $184.2\times$ latency \\
\hline
\end{tabular}
\end{table}

We then inject two branch faults under all three configurations: one gibberish branch in the kernel-submission code and one sponge branch in the compiled-cache lookup. The gibberish fault yields PPL ratios of $35{,}831.1\times$, $35{,}831.2\times$, and $37{,}428.9\times$ for the unprotected, FaR, and LM-Fix configurations, respectively; all three runs keep the process alive and return visibly gibberish text. The sponge fault yields Latency ratios of $188.0\times$, $206.1\times$, and $184.2\times$, respectively, while preserving the generated text. Neither FaR nor LM-Fix meaningfully mitigates the attack effects caused by branch faults in code introduced by JIT compilation.

The reason is structural. FaR rewires selected model parameters, while LM-Fix monitors and restores corrupted weight state. Our attack instead faults conditional branches in the code introduced by JIT compilation and leaves model weights intact. The resulting runtime control-flow deviation therefore falls outside both defenses' protection scope. FaR and LM-Fix exemplify current weight-level defenses: they resist faults in model weights but do not protect the JIT-introduced control flow targeted by our branch faults.

\subsection{End-to-End Rowhammer Attacks}
\label{sec:eval_attack}
We finally evaluate their physical exploitability with Rowhammer. We match vulnerable branch bits to profiled DRAM bit-flips, place the corresponding code pages on compatible physical pages using memory waylaying, and induce the required single-bit faults in CPU DRAM while inference executes on the GPU. \Cref{tab:gibberish_disruption} and \Cref{tab:sponge_disruption} report the two attack effects.

\begin{table}[!t]
\centering
\caption{Gibberish-output results from end-to-end Rowhammer attacks.}
\label{tab:gibberish_disruption}
\footnotesize
\setlength{\tabcolsep}{3pt}
\begin{tabular}{l c c c}
\hline
Model & Clean PPL & Faulted PPL & PPL ratio \\
\hline
\multicolumn{4}{l}{\textit{Text-only inference: WikiText-2}} \\
Qwen3 & 8.78 & 2{,}523.61 & $287.3\times$ \\
DeepSeek-R1 & 24.12 & $1.20{\times}10^{8}$ & $4.98{\times}10^{6}\times$ \\
\hline
\multicolumn{4}{l}{\textit{Multimodal inference: correct-choice PPL}} \\
Gemma3 & 56.92 & $4.11{\times}10^{8}$ & $7.23{\times}10^{6}\times$ \\
Qwen2.5-VL & 1.04 & $1.46{\times}10^{4}$ & $1.40{\times}10^{4}\times$ \\
\hline
\end{tabular}
\end{table}

\begin{table}[!t]
\centering
\caption{Latency-amplification results from end-to-end Rowhammer attacks.}
\label{tab:sponge_disruption}
\footnotesize
\setlength{\tabcolsep}{2.5pt}
\begin{tabular}{l c c c}
\hline
Model & Clean latency (s) & Faulted latency (s) & Latency ratio \\
\hline
Qwen3 & 0.847 & 105.890 & $124.97\times$ \\
DeepSeek-R1 & 0.785 & 76.466 & $97.38\times$ \\
Gemma3 & 0.732 & 77.916 & $106.39\times$ \\
Qwen2.5-VL & 1.046 & 107.270 & $102.60\times$ \\
\hline
\end{tabular}
\end{table}

As shown in \Cref{tab:gibberish_disruption}, a single vulnerable branch-bit fault increases PPL across all four models while the inference process remains alive and returns a completed response. The PPL ratios range from $287.3\times$ to $7.23{\times}10^{6}$. For Qwen3 and DeepSeek-R1, the corresponding 32-token generation runs replace coherent text with gibberish, including repeated numeric fragments such as \texttt{0.0.1.0} and repeated occurrences of \texttt{the}. For the multimodal models, we aggregate the correct-choice PPL by averaging the three question-level NLLs before converting the result to PPL. The raw generations are visibly corrupted across all three questions. Consequently, the generated outputs become unusable gibberish across both text-only and multimodal models. Thus, the attack depletes inference without triggering a conventional process failure.

As shown in \Cref{tab:sponge_disruption}, the generated token sequence remains identical to clean inference, while request latency increases from 0.73--1.05\,s to 76.47--107.27\,s, corresponding to a \(97.38\times\)--\(124.97\times\) slowdown. This demonstrates a \emph{correct-output sponge attack}: the server performs significantly more work for the same request without changing the returned output. Unlike input-driven LLM sponge attacks~\cite{gao2024verbose,dong2025engorgio,li2026loopllm}, our fault-induced sponge preserves the clean output at the token level while significantly increasing the serving latency.

Overall, these demonstrated attacks show that a single Rowhammer-induced fault in CPU-resident JIT control code can propagate across the CPU--GPU boundary and disrupt GPU-executed LLM inference without direct access to GPU memory. Across all four models, the attacks present both gibberish-output and correct-output sponge effects while preserving process liveness.

\section{Discussion}

\subsection{Mitigations}
\name can be mitigated by detecting anomalous inference behavior, and protecting JIT-introduced runtime control flow.

Output-level checks such as redundant inference, output sanity checks, and recomputation may detect gibberish or other output-corruption effects. In contrast, the correct-output sponge preserves the model output and therefore requires request-latency monitoring or lightweight JIT runtime telemetry, such as unexpected recompilation, fallback execution, or kernel-submission activity. However, the enforcement logic of these checks still depends on runtime control flow. Prior Rowhammer work on hardened cryptographic implementations has shown that such enforcement branches can themselves become fault targets when they remain in the same unprotected code path~\cite{liang2025achilles}.

A complementary software-side mitigation is therefore to harden high-impact runtime branches, for example through duplicated conditional jumps, branch-outcome checks, secure enclaves, code-page isolation, or code-integrity mechanisms~\cite{barry2016compilation,chen2017camfas,costan2016intel,gruss2018another}. Since our faults redirect execution to an existing successor in the program's control-flow graph, protection should cover branch outcomes or code-page integrity rather than only illegal control transfers.

\subsection{Limitations}
\mypara{Decision-guided fault-vulnerable code analysis}
Our current implementation focuses on PyTorch 2.9.1. Updates to the JIT stack may change source routines, and branch locations. Also, other JIT frameworks such as JAX/XLA and PJRT have different routines and branches.
Given \name anchors its analysis on inference-critical JIT semantics, applying it to a new version or build or a different framework requires rerunning the source localization, LLVM IR analysis, and source-to-binary fault mapping.
We leave such cross-framework and cross-backend validation to future work.

\mypara{Rowhammer attacks}
Our end-to-end attacks are evaluated on one non-ECC DDR4 platform and we do not evaluate ECC-protected DDR4 and newer DDR5 systems. While Error-Correcting Code (ECC) and Target Row Refresh (TRR) have been deployed by industry to mitigate Rowhammer, recent work has demonstrated practical bypasses~\cite{cojocar2019exploiting,kamadan2025ecc,meyer2026phoenix}, suggesting that \name could also be instantiated on newer platforms. We leave such validation to future work.

\section{Conclusion}
We presented \name, a model-agnostic single-bit Rowhammer attack against the host-side JIT serving control plane of LLM inference. \name uses decision-guided analysis to identify vulnerable branches governing compiled-artifact selection and GPU-work submission, and demonstrates that a single CPU-DRAM fault can propagate to GPU-executed inference, causing either gibberish output or significant correct-output latency amplification while preserving process liveness.

\bibliographystyle{plain}
\bibliography{bib/defs,bib/references}

\end{document}